\documentclass[twocolumn,prb]{revtex4}
\usepackage{amsmath,amsbsy,amssymb,graphicx,color,pifont,tikz}
\usepackage[mathscr]{euscript}
\usepackage{fouriernc}

\def\babc{\begin{subequations}}
\def\eabc{\end{subequations}}
\def\be{\begin{equation}}
\def\ee{\end{equation}}
\def\ba{\begin{array}}
\def\ea{\end{array}}
\def\nn{\nonumber}
\def\hh{\hspace*{0.2mm}}
\def\h{\hspace*{0.5mm}}
\def\5{\hspace*{5mm}}
\def\lgr{\left\lgroup\hspace*{-1.25mm}}
\def\rgr{\hspace*{-1.25mm}\right\rgroup}

\def\k{\kappa}
\def\bk{\boldsymbol{\kappa}}
\def\bs{\boldsymbol{\sigma}}
\def\2{{\textstyle{\frac12}}}
\def\4{{\textstyle{\frac14}}}
\def\bh{\boldsymbol{h}}

\def\boa{\boldsymbol{a}}
\def\bob{\boldsymbol{b}}

\begin{document}

\title{Group Structure of Wilson Loops in 2D Models with 2- and 4-Band Energy Spectra}

\author{T. Supatashvili$^1$, M. Eliashvili$^{1,2}$, and G. Tsitsishvili$^{1,2}$}
\affiliation{
$^{1}$Department of Physics, Tbilisi State University, Chavchavadze Ave. 3, Tbilisi 0179, Georgia\\
$^{2}$Razmadze Mathematical Institute, Tbilisi State University, Tamarashvili Str. 6, Tbilisi 0177, Georgia}

\begin{abstract}
We consider a tight-binding model defined by a matrix Hamiltonian over 2D Brillouin zone.
Multiband energy spectrum gives rise to a non-Abelian gauge structure set by the Berry connections.
The corresponding curvature $F_{\mu\nu}$ vanishes throughout the Brillouin zone except an isolated
points where $F_{\mu\nu}$ is singular. Combining the singular behaviour of $F_{\mu\nu}$ with
non-Abelian Stokes theorem allows to avoid the path ordering procedure in studying the structure
of Wilson loops. 2D models with 2-band and 4-band energy spectra are considered as a demonstrative
examples and the group structure of the corresponding Wilson loops is revealed.
\end{abstract}

\maketitle

\section*{1. Introduction}

Appearance of topological insulators \cite{topins} entailed the concept of topological phases implying the novel states of matter
that are specified in terms of topological structures rather than by the notion of symmetries. The simplest construction producing
topological phases is the 1D tight-binding model of polyacetylene known as the Su-Schrieffer-Heeger model.\cite{ssh}
Variety of different models exhibiting nontrivial topological phases in 2 and 3 spatial dimensions have been extensively studied over the past decade.\cite{model1,model2,model3} Topological indices which label such phases of tight binding models are usually expressed as
integrals of the Berry connections\cite{berry} over the Brillouin zone (BZ).

On the other hand, the models endowed with multiband energy spectra give rise to non-Abelian gauge structures\cite{vanderbilt}
in the BZ. The standard construction containing topological features of a non-Abelian gauge field is the Wilson loop (WL)
considered as an effective tool for describing topological indices in multiband models.\cite{dai,wang,cano,bradlyn,bouhon,guo,wieder}
Practical calculations of WLs become complicated by the path-ordering procedure. In our case extra obstacles occur due to
non-triviality of fundamental group of a torus $\pi_1(T^2)=\mathbb Z\times\mathbb Z$.

The key observation of the present note is that non-Abelian Berry connection built up via the multiband state vectors
represents pure gauge but with point-like singularities in the BZ. Then the curvature tensor $F_{\mu\nu}$ identically
vanishes except the aforementioned points where $F_{\mu\nu}$ is singular. We then employ the non-Abelian Stokes
theorem\cite{nast} expressing WL in terms of a surface integral of $F_{\mu\nu}$. Provided the later
is non-vanishing only at certain isolated points we manage to fully benefit from the non-Abelian Stokes theorem so that
the WL related to an isolated singular point is expressed in the closed form in terms of Berry phases.

Singular points turn out to be of two kinds. One results from a gap closure, while the other occurs in gapped states.
In the present note we discuss the gapped states only, for which the Berry phases are well defined, while for the gapless
states one encounters the poorly understood problem of level crossing.

For the gapped states we argue that the WL around an isolated singular point is trivial (identity matrix).
Employing this fact we show that the set of WLs along the elements of $\pi_1(T^2)$ reproduce the group
structure of $\pi_1(T^2)$.

In Sect. 2 we present the non-Abelian gauge structure as resulting from the multiband energy spectra and point out
the key observation of the given account implying that the curvature $F_{\mu\nu}$ vanishes throughout the BZ except
an isolated points where $F_{\mu\nu}$ is singular. In Sect. 3 we discuss the fundamental group of a torus.
In Sect. 4 we merge the point-like singular behaviour of $F_{\mu\nu}$ with non-Abelian Stokes theorem and derive
the simple analytic expression relating the WL for an isolated singular point to the Berry phases. In Sect. 5 we consider
the 2-band model and trace out the group structure of the corresponding WLs. In Sect. 6 we comment on the 4-band model.
Sect. 7 is devoted for summary.

\section*{2. Non-Abelian Berry Structure}

Let $H(\bk)$ be an $N\times N$ hermitian matrix Hamiltonian defined over 2D BZ with
$H(\k_1+2\pi,\k_2)=H(\k_1,\k_2+2\pi)=H(\k_1,\k_2)$.
Let $E_n(\bk)$ and $\psi_n(\bk)$ be its eigenvalues and eigenstates
\be
H(\bk)\hh\psi_n(\bk)=E_n(\bk)\hh\psi_n(\bk)
\ee
where $n=1,\ldots,N$. The eigenstates satisfy the orthogonality and completeness relations
\babc
\begin{align}
&\sum_p(\psi^\dag\hspace*{-1.3mm}_m)_p(\psi_n)_p=\delta_{mn},\\
&\sum_n(\psi_n)_p(\psi^\dag\hspace*{-1.3mm}_n)_q=\delta_{pq}.
\end{align}
\eabc

Introduce the matrix-valued connection
\be
(A_\mu)_{mn}=i\hspace*{0.3mm}\psi^\dag\hspace*{-1.4mm}_n\partial_\mu\psi_m,
\hspace*{10mm}
\partial_\mu=\partial/\partial\k_\mu,
\ee
and trace out how $A_\mu$ responds to the unitary transformation
\be
H(\bk)\to\tilde{H}(\bk)=\Omega^\dag(\bk)H(\bk)\hh\Omega(\bk)
\ee
where $\Omega(\bk)$ is double-periodic unitary matrix, so that $\tilde{H}(\bk)$ inherits the periodicity of $H(\bk)$.

Consider the corresponding eigenvalue problem
\be
\tilde{H}(\bk)\hh\tilde\psi_n(\bk)=E_n(\bk)\hh\tilde\psi_n(\bk).
\ee
Relation between $\psi_n$ and $\tilde\psi_n$ appears as
\be
(\tilde\psi_m)_p=(\Omega^\dag)_{pq}(\psi_m)_q
\ee
and can be rewritten as
\be
(\tilde\psi_m)_p=(U^\dag)_{mn}(\psi_n)_p
\ee
where the unitary matrix $U(\bk)$ is given by
\be
U_{mn}=\psi^\dag\hspace*{-1.3mm}_n\Omega\psi_m.
\ee

Introduce the transformed connection
\be
(\tilde{A}_\mu)_{mn}=i\hspace*{0.3mm}\tilde\psi^\dag\hspace*{-1.4mm}_n\partial_\mu\tilde\psi_m.
\ee
Substituting (7) into (9) we find
\be
\tilde{A}_\mu=U^\dag A_\mu U-iU^\dag\partial_\mu U
\ee
where from it is evident that the Berry connections (3) constitute the non-Abelian gauge field defined over $\bk$-space.

Expressing the corresponding curvature tensor
\be
F_{\mu\nu}=\partial_\mu A_\nu-\partial_\nu A_\mu+i\hspace*{0.3mm}[A_\mu,A_\nu]
\ee
in terms of the connections (3) we obtain the central point of our consideration
\be
(F_{\mu\nu})_{mn}=i\hspace*{0.3mm}\psi^\dag\hspace*{-1.4mm}_n(\partial_\mu\partial_\nu-\partial_\nu\partial_\mu)\hh\psi_m.
\ee
This expression implies, that $F_{\mu\nu}=0$ everywhere in the Brillouin zone except the points where  $\psi_n(\bk)$ are singular.
Using the standard terminology, the non-Abelian Berry connections set by (3) represent pure gauge with point-like singularities.
In general one may have several singular points within the BZ.

Such singularities can be specified by a Wilson loop (WL)
\be
W_\gamma=P\h{\rm exp}\h\Bigg(-i\oint_\gamma\boldsymbol{A}\hh d\boldsymbol{l}\Bigg)
\ee
where the integration is implied along the closed path $\gamma\in{\rm BZ}$, and $P$ stands for path-ordering.

Provided the BZ is topologically equivalent to a 2D torus, the set of integration closed curves is non-trivial.
This feature imposes specific constraints on the structure of WLs. At this point some comments
on the fundamental group of a torus are in order.

\section*{3. Fundamental Group $\pi_1(T^2)$}

Fundamental group of a 2D torus is $\pi_1(T^2)=\mathbb Z\times\mathbb Z$, {\it i.e.} each loop can be specified by
a pair of integers $(m,n)$, where $m$ and $n$ count windings over two principal circle of a torus.
These closed curves are parameterized by the variables $\k_1$ and $\k_2$.
In Fig. 1 we depict several particular cases of loops with different values of $(m,n)$.

\begin{figure}
\begin{center}
\includegraphics{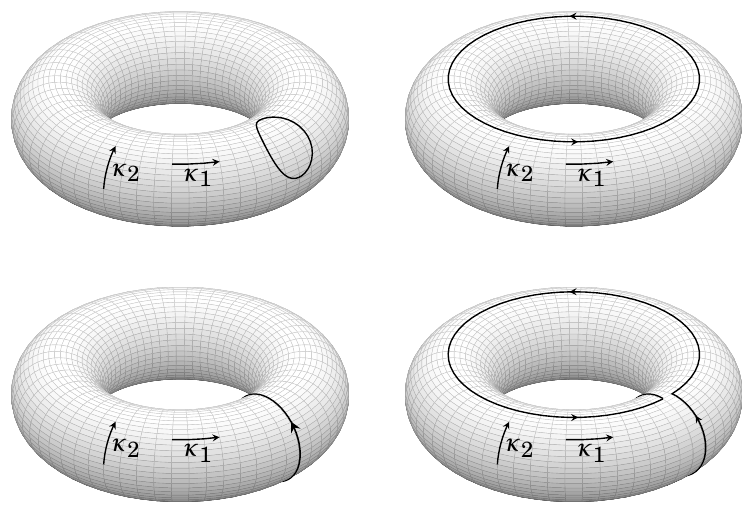}
\end{center}
\caption{{\bf Top left}: Loop contractible to a point $(m,n)=(0,0)$.
{\bf Top right}: $(m,n)=(1,0)$. {\bf Bottom left}: $(m,n)=(0,1)$. {\bf Bottom right}: $(m,n)=(1,1)$.}
\end{figure}

Defining the homotopy group $\pi_1(M)$ of a manifold $M$ we operate with closed curves.
In order to define the composition of two curves (group operation) it is necessary to endow
each curve by a point $x\in M$ where it starts and ends. This point which is usually referred
to as the base-point, must be one and the same for all curves under consideration otherwise the
composition cannot be defined. Correspondingly, the homotopy group is denoted by $\pi_1(M,x)$.
The group $\pi_1(M,x')$ with $x'\ne x$ is isomorphic to $\pi_1(M,x)$.

In this light we consider the WLs along the closed curves with a given base-point $\bk_B\in{\rm BZ}$.
Relation between the WLs with different $\bk_B$ is discussed in the end.

\section*{4. Non-Abelian Stokes Theorem}

Non-Abelian Stokes theorem relates the WL to the surface integral of the curvature $F_{\mu\nu}$.
It appears as\cite{nast,nast1}
\be
P\h{\rm exp}\h\Bigg(\hspace*{-0.25mm}-i\oint_{\partial S}\hspace*{-1.5mm}\boldsymbol{A}d\bk\Bigg)
=\mathscr P\h{\rm exp}\Bigg(\hspace*{-0.25mm}-i\int_S w^\dag F_{\mu\nu}\hh w\h dS_{\mu\nu}\Bigg)
\ee
where $w(\bk)$ is some unitary matrix, $\mathscr P$ implies certain ordering procedure (different from $P$) and
$dS_{\mu\nu}=\2\epsilon_{\mu\nu}d\k_1d\k_2$. Definitions for $w(\bk)$ and $\mathscr P$ can be found {\it e.g.}
in Ref. [16], while for our purposes no need in these details.

Let $\{\bk_1,\bk_2,\ldots,\bk_L\}$ be the set of point where $F_{\mu\nu}$ is singular, and consider the loop
$\gamma$ shown in Fig. 2 comprising no singular points in its interior. We do not restrict the surface bounded
by $\gamma$ to be necessarily within a single BZ; the surface being simply connected may cover parts of neighbouring BZs.
\begin{figure}[h]
\begin{center}
\includegraphics{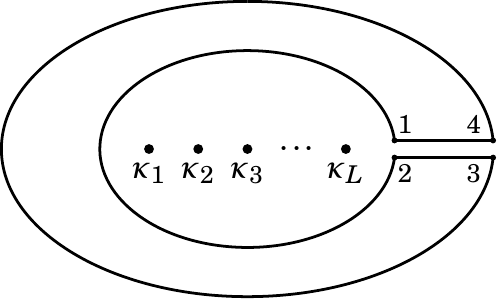}
\end{center}
\caption{Closed curve $1\to2\to3\to4\to1$ comprises no singularities, hence the corresponding Wilson loop is trivial $W_\gamma=\mathbb1$.}
\end{figure}

Provided $F_{\mu\nu}=0$ throughout the surface bounded by $\gamma$, the non-Abelian Stokes theorem (14) implies
the triviality of the WL, {\it i.e.} $W_\gamma=\mathbb1$.

Segmenting the loop $\gamma$ we write
\be
W_{14}W_{43}W_{32}W_{21}=\mathbb1
\ee
where $W_{ba}$ is the contribution from a segment $a\to b$.

Using $U\equiv W_{14}=[W_{32}]^{-1}$ we rewrite (15) as
\be
W(\Gamma)=U^\dag W(\gamma)U
\ee
where $\Gamma$ and $\gamma$ denote the loops $4\to3$ and $1\to2$ respectively,
encircling one and the same set of singular points.

Summarising, for the non-Abelian Berry connection representing pure gauge with point-like singularities
we arrive to the following conclusion: if two distinct loops $\Gamma$ and $\gamma$ enclose one
and the same set of singular points, then the corresponding WLs are unitary equivalent.

\section*{\it 4.1. Isolated Singularity}

Consider the relation (16) with the loops $\Gamma$ and $\gamma$ encircling one isolated singular point $\bk_0$,
and squeeze $\gamma$ to a point $\bk_0$, {\it i.e.}
\be
W_\gamma=P\h{\rm exp}\h\Bigg(-i\oint_\gamma\boldsymbol{A}\hh d\bk\Bigg)
\ee
where the integration path $\gamma$ is set by $\bk=\bk_0+\varepsilon(\cos\alpha,\sin\alpha)$
with $0\leqslant\alpha\leqslant2\pi$ and $\varepsilon\to0$.

In order to calculate (17) we turn back to the non-Abelian Stokes formula (14).
In this case the integration surface ($S$) in its right hand side is the area to be shrunk to zero,
but capturing the point $\bk_0$. Therefore the only contribution to the surface integral comes
from the point $\bk_0$, meaning that the ordering $\mathscr P$ is no longer relevant. We then have
\begin{align}
W_\gamma
&={\rm exp}\hh\Bigg(\hspace*{-0.25mm}-i\int_{S\to0}\hspace*{-4.0mm}w^\dag(\k)F_{12}(\k)w(\k)\hh d\k_1d\k_2\Bigg)=
\nn\\
&=w^\dag(\k_0)\hh{\rm exp}\hh\Bigg(\hspace*{-0.25mm}-i\int_{S\to0}\hspace*{-4.0mm}F_{12}(\k)\hh d\k_1d\k_2\Bigg)w(\k_0)
\end{align}

Introduce the flux matrix
\be
\Phi(\bk_0)=\int_{S\to0}\hspace*{-4.0mm}F_{12}(\k)\hh d\k_1d\k_2.
\ee
Contribution to the right hand side of (19) originates from the singularity of $F_{12}=\partial_1A_2-\partial_2A_1+i[A_1,A_2]$,
which comes from the Abelian part $\partial_1A_2-\partial_2A_1$, but not from the commutator $[A_1,A_2]$.
Therefore, the essential part of (19) is given by
\be
\Phi(\bk_0)=\int_{S\to0}\hspace*{-4.0mm}(\partial_1A_2-\partial_2A_1)\hh d\k_1d\k_2
=\oint_\gamma A_\mu\h d\k_\mu
\ee
where the contour $\gamma$ is the same as in (17). We thus arrive to
\be
W_\gamma=w^\dag(\bk_0)\hh e^{-i\hspace*{0.1mm}\Phi(\bk_0)}w(\bk_0)
\ee
Substituting (21) into (16) we find
\be
W_\Gamma=[w(\bk_0)U]^\dag e^{-i\hspace*{0.1mm}\Phi(\bk_0)}[w(\bk_0)U]
\ee
{\it i.e.} if $\Gamma$ is an arbitrary loop enclosing an isolated singular point $\bk_0$ then
$W(\Gamma)$ is unitary equivalent to the matrix $e^{-i\hspace*{0.1mm}\Phi(\bk_0)}$.

In the following sections we use the result (22) to study the structure of WLs for a general 2-band model in 2D.

\section*{5. Two-Band Model}

Consider $2\times2$ matrix Hamiltonian
\be
H=\bh(\bk)\cdot\bs
\ee
where $\bh=(h_1,h_2,h_3)$ is defined on the BZ $\bk=(\k_1,\k_2)\in T^2$ (2D torus),
and $\bs$ are the Pauli matrices.

Irrespectively of the explicit form of the functions $h_a(\bk)$ the Hamiltonian (23) exhibits the symmetry
\be
\sigma_2\hh H^*\hspace*{-0.1mm}(\bk)\hh\sigma_2=-\hh H(\bk)
\ee
leading to the symmetric spectrum $E_\pm(\bk)=\pm|\bh(\bk)|$. The corresponding eigenstates look as ($h\equiv|\bh|$)
\babc
\begin{align}
\psi_+&=\frac{1}{\sqrt{2h(h-h_3)}}\lgr\ba{c}h_1-ih_2\vspace*{1.45mm}\\+h-h_3\ea\rgr\\
\psi_-&=\frac{1}{\sqrt{2h(h-h_3)}}\lgr\ba{c}-h+h_3\vspace*{1.45mm}\\h_1+ih_2\ea\rgr
\end{align}
\eabc
with $\psi_-=-i\sigma_2\psi_+\hspace*{-1.7mm}^*$ as it follows from (24).

Substituting (25) into (3) the Berry connections appear as
\babc
\begin{align}
(A_\mu)_{\pm\pm}&=\pm\h\frac{h_1\partial_\mu h_2-h_2\partial_\mu h_1}{2h(h-h_3)}\\
(A_\mu)_{-+}&=\frac{i\partial_\mu(h_1+ih_2)}{2h}-\frac{i(h_1+ih_2)\partial_\mu(h-h_3)}{2h(h-h_3)}
\end{align}
\eabc
Note that $A_\mu$ is a hermitian matrix, {\it i.e.} $(A_\mu)_{+-}=[(A_\mu)_{-+}]^*$.

We distinguish between two different kinds of singularities. One occurs for $h_3=h\ne0$, and is not related to a gap closing.
The other occurring for $h=0$ represents the gap closure. In the given paper we discuss the gapped states only, for which
the Berry phase is well defined, while for gapless states one encounters the poorly understood problem of level crossing.

\section*{\it 5.1. Wilson Loops for $(m,n)=(0,0)$}

We start by the case of a loop with an isolated singular point $\bk_0$, {\it i.e.} we calculate (22).

Let $\bk_0$ be the point where $h_1=h_2=0$ and $h_3>0$. Taking $\bk=\bk_0+\varepsilon(\cos\alpha,\sin\alpha)$
with $\varepsilon\to0$ we have $h_1+ih_2=\mathscr O(\varepsilon^p)$ and $h-h_3=\mathscr O(\varepsilon^{2p})$ where $p>0$.
Using these in (26b) we find $(A_\mu)_{-+}d\k_\mu=\mathscr O(\varepsilon^p)$ leading to $\Phi_{\pm\mp}(\bk_0)=0$
as $\varepsilon\to0$.

In order to calculate $(A_\mu)_{\pm\pm}$ we use the polar coordinates in $h$-space and introduce the unit vector
\babc
\begin{align}
\hat h_1&=h_1/h=\sin\theta(\bk)\cos\phi(\bk),\\
\hat h_2&=h_2/h=\sin\theta(\bk)\sin\phi(\bk),\\
\hat h_3&=h_3/h=\cos\theta(\bk).
\end{align}
\eabc
Then the diagonal components (26a) appear as
\be
(A_\mu)_{\pm\pm}=\pm\big[\partial_\mu\phi(\bk)\big]\cos^2\big[\2\theta(\bk)\big].
\ee

As a matter of $h_1(\bk_0)=h_2(\bk_0)=0$ and $h_3(\bk_0)>0$ we have $\theta(\bk_0)=0$.
Therefore, integrating around $\bk_0$ we obtain
\be
\Phi_{\pm\pm}(\bk_0)=\oint_{\varepsilon\to0}\hspace*{-3mm}(A_\mu)_{\pm\pm}\hh d\k_\mu
=\pm\oint_{\varepsilon\to0}\hspace*{-3mm}(\partial_\mu\phi)\hh d\k_\mu
=\pm2\pi\mathbb Z.
\ee
As a result we arrive to $\Phi(\bk_0)=2\pi\mathbb Z\sigma_z$, hence $e^{-i\Phi(\bk_0)}=\mathbb1$.
Using this in (22) we come to
\be
W_\Gamma=\mathbb1.
\ee

The same result can be obtained directly from (17) {\it i.e.} without applying the steps (18) -- (22).
As we already pointed out, the matrix $A_\mu d\k_\mu$ becomes diagonal as $\varepsilon\to0$.
Then the ordering is no longer relevant in (17) which by use of (29) leads to $W_\gamma=\mathbb1$.
Using the latter in (16) we reproduce (30).

Consider the case of the integration loop encircling several singular points.
We first comment on the particular case of the loop $\gamma$ shown in Fig. 3a and encircling three singular points.

\begin{figure}[h]
\vspace*{1mm}
\begin{center}
\includegraphics{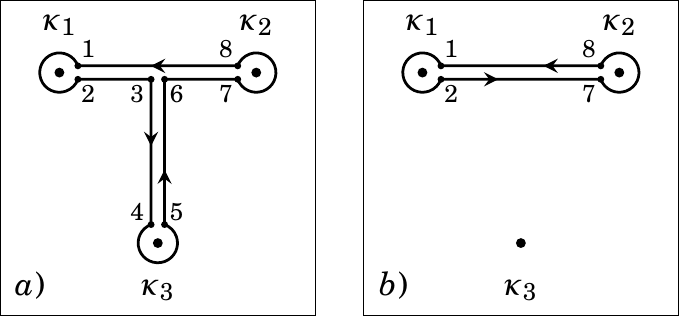}
\end{center}
\caption{{\bf Left}: Contour $\gamma$ enclosing three singular points $\bk_{1,2,3}$.
{\bf Right}: Contour $\gamma'$ enclosing two singular points $\bk_{1,2}$.}
\end{figure}

For the corresponding WL we have
\be
W_\gamma=W_{18}W_{87}W_{76}W_{65}W_{54}W_{43}W_{32}W_{21}
\ee
where $W_{ba}$ denotes the contribution from the segment $a\to b$.

Consider the segment $3\to4\to5\to6$. The corresponding Wilson line is
$W_{65}W_{54}W_{43}$ where $W_{65}=[W_{43}]^{-1}$, and $W_{54}$
is the WL around $\bk_3$ which is trivial. Then $W_{65}W_{54}W_{43}$ is also trivial,
and (31) takes the form
\be
W_\gamma=W_{18}W_{87}W_{76}W_{32}W_{21}=W_{18}W_{87}W_{72}W_{21}
\ee
{\it i.e.} the two integration loops $\gamma$ and $\gamma'$ shown in Fig. 3 produce one and the same WL.

Applying the same arguments to the path $2\to7\to8\to1$ in Fig. 3b we find
$W_{18}W_{87}W_{72}=\mathbb1$. We thus obtain $W_\gamma=W_{21}$ where $W_{21}$
is the WL enclosing $\bk_1$ which is trivial.

Remind that the WL along an arbitrary loop $\Gamma$ enclosing the same $\bk_{1,2,3}$ is unitary equivalent
to $W_\gamma$ {\it i.e.} $W_\Gamma=U^\dag W_\gamma U$. Consequently, provided $W_\gamma=\mathbb1$
we have $W_\Gamma=\mathbb1$ for all $\Gamma$.

In the case of an arbitrary amount of the enclosed singular points, the segments like $3\to4\to5\to6$ in Fig. 3a
can be resected step by step until the integration curve is reduced to the one enclosing an isolated singular point.
The latter yields trivial WL. Therefore, irrespectively of the amount of the enclosed singular points all WLs with
$(m,n)=(0,0)$ are trivial.
Mind that this occurs due to the fact that the Berry connection is a pure gauge with point-like singularities.

\section*{\it 5.2. Wilson Loops for $(m\ne0,0)$}

We first comment on the case of $(m,n)=(1,0)$. Consider the closed contour $a\to b\to a'\to c\to a$ shown in Fig. 4a.

\begin{figure}[h]
\begin{center}
\includegraphics{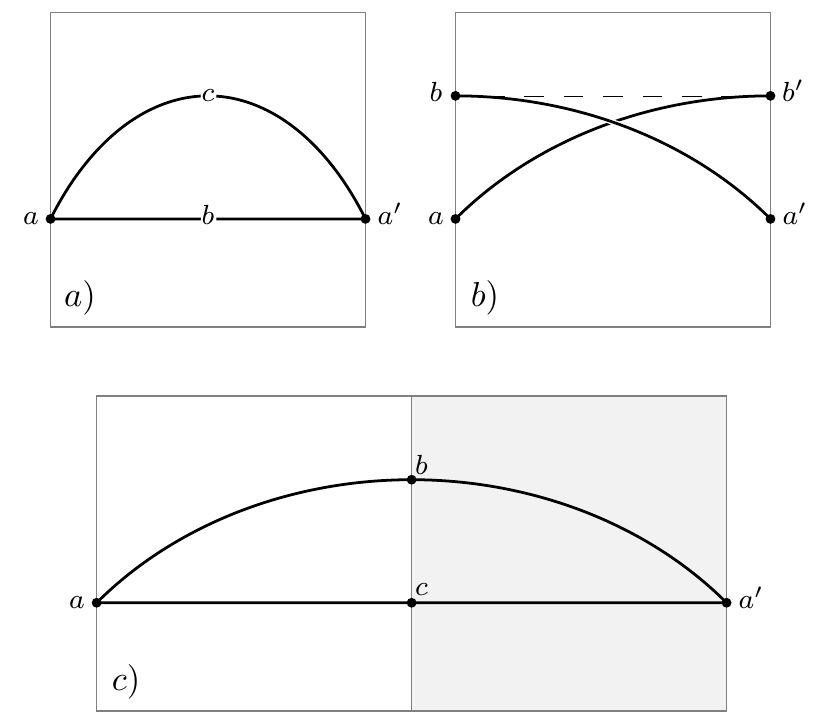}
\end{center}
\caption{{\bf Top left}: WLs along the paths $a\to b\to a'$ and $a\to c\to a'$ are identical.
{\bf Top right}: The curve $a\to b'(=b)\to a'$ belonging to $(2,0)$-class of homotopy.
{\bf Bottom}: The curve from top right panel redrawn across two BZs.}
\end{figure}

According to our previous results the corresponding WL is an identity irrespectively of the amount of singularities enclosed
\be
\mathbb1=W_{a\to b\to a'\to c\to a}=W_{a'\to c\to a}W_{a\to b\to a'}
\ee
where from we find
\be
W_{a\to c\to a'}=W_{a\to b\to a'}
\ee
signifying that the WL is insensitive to the trajectory of traveling from $a$ to $a'$. Therefore all WLs with $(1,0)$ can be
presented by integrating along the straight lines in the BZ.

Consider the case of $(2,0)$. This is depicted in Fig. 4b and can be redrawn over two BZs as
$a\to b\to a'$ in Fig. 4c.

Remark that the previously discussed loop shown in Fig. 2 is not necessarily within a single BZ, but may traverse through
several BZs, like the one in Fig. 4c. Therefore the WL for the closed path in Fig. 4c is an identity. Consequently
\be
W_{a\to b\to a'}=W_{a\to c\to a'}
\ee
implying that the WL for $a\to a'$ is independent of the shape of the curve connecting $a$ to $a'$.
Moreover, the WL in the right-hand side of (35) can be rewritten as $W_{c\to a'}W_{a\to c}$ where the periodicity
implies $W_{c\to a'}=W_{a\to c}$. Summarizing, for $(2,0)$ we have $W_{a\to a'}=(W_{a\to c})^2$.
Generalization to $(m,0)$ is straightforward and can be expressed as
\be
W(m,0)=[W(1,0)]^m
\ee
where $W(m,n)$ will be used hereafter for the WL corresponding to a closed path belonging to the $(m,n)$-class
of homotopy of a torus.

In the same way we handle with the case of $(0,n)$ and obtain
\be
W(0,n)=[W(0,1)]^n.
\ee

WLs considered in this section though being non-trivial do not depend on the shape
of the integration curves. Therefore we ignore the latter as an argument.

\section*{\it 5.3. Wilson Loops for $(m\ne0,n\ne0)$}

Consider the path with $(m,n)=(1,2)$ shown in Fig. 5a.
In the expanded form this can be redrawn as $a\to c'$ in Fig. 5b.

\begin{figure}[h]
\begin{center}
\includegraphics{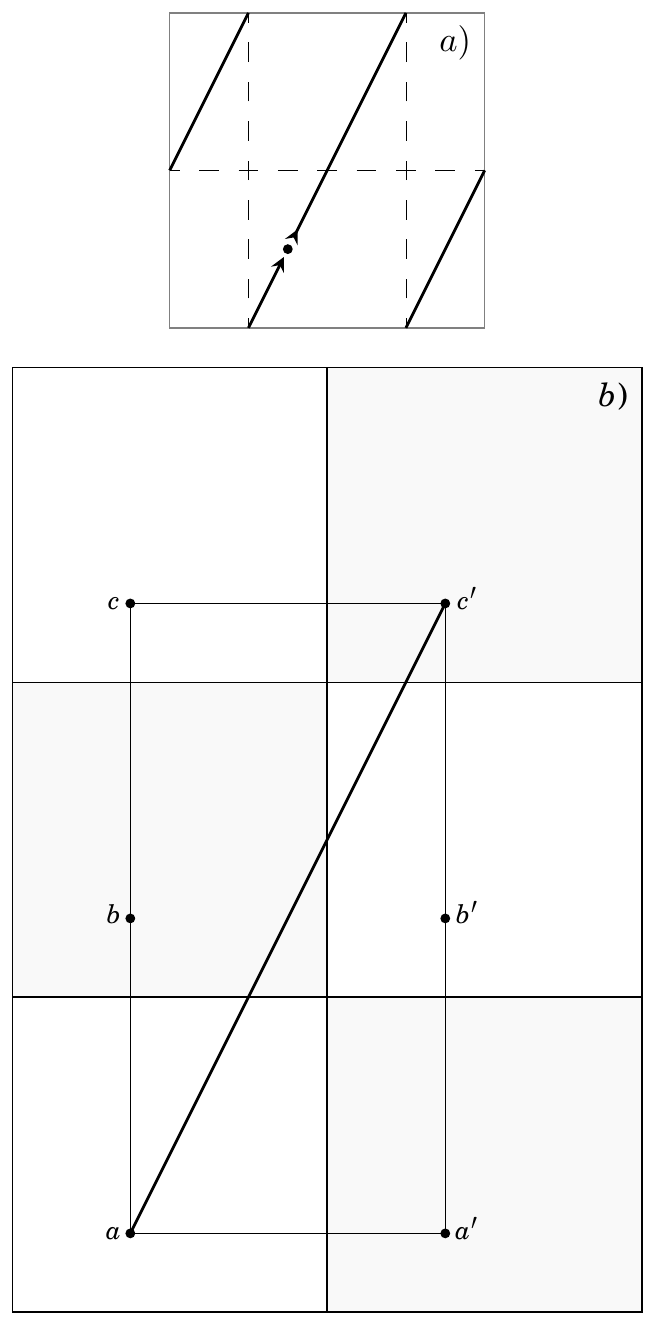}
\end{center}
\caption{{\bf Top}: The curve $(1,2)$ drawn within a single BZ. {\bf Bottom}: The same curve drawn in extended way over several BZs.}
\end{figure}

In according to our previous results, the WL for the closed path $a\to a'\to b'\to c'\to a$
is an identity matrix irrespectively of the amount of singularities enclosed. We then have
\be
W_{a\to c'}=W_{b'\to c'}W_{a'\to b'}W_{a\to a'}.
\ee
We have $W_{a\to a'}=W(1,0)$ and $W_{a'\to b'}=W_{b'\to c'}=W(0,1)$.
In this light we come to
\be
W(1,2)=[W(0,1)]^2W(1,0).
\ee
Performing the same with $a\to b\to c\to c'\to a$ we come to
\be
W(1,2)=W(1,0)[W(0,1)]^2.
\ee
These observations are trivially extendable to the general case of $(m,n)$ leading to
\begin{align}
W(m,n)&=[W(1,0)]^m[W(0,1)]^n=\nn\\
&=[W(0,1)]^n[W(1,0)]^m
\end{align}
hence the matrices $W(m,n)$ all commute among each other.

For the negative values of $m$ and $n$ the relation (41) is valid in the sense that $W(-1,0)$ is
the inverse of $W(1,0)$, and so on.

\section*{\it 5.4. Group Structure of Wilson Loops}

Summarising the preceding sections, the matrices $W(m,n)$ form the Abelian group with the multiplication law
\be
W(m,n)W(k,l)=W(m+k,n+l).
\ee
In this light the matrices $W(m,n)$ produce the unitary representation of the homotopy group $\pi_1(T^2)$.

Provided the matrices $W(m,n)$ all commute among themselves,
they can be simultaneously diagonalized by some unitary transformation $U$.
For the particular choice of the overall phases in (25) we have ${\rm Tr}(A_\mu)=0$.
This leads to ${\rm det}(W)=1$ which together this with $W^\dag W=\mathbb1$ implies
the eigenvalues of $W$ are unimodular. Therefore we have
\be
U^\dag W(m,n)U=\lgr\ba{cc}e^{+i(m\xi+n\zeta)}&0\vspace*{2mm}\\0&e^{-i(m\xi+n\zeta)}\ea\rgr
\ee
where $\xi$ and $\zeta$ are set by
\babc
\begin{align}
U^\dag W(1,0)U&=\lgr\ba{cc}e^{+i\xi}&0\vspace*{2mm}\\0&e^{-i\xi}\ea\rgr,\\
U^\dag W(0,1)U&=\lgr\ba{cc}e^{+i\zeta}&0\vspace*{2mm}\\0&e^{-i\zeta}\ea\rgr.
\end{align}
\eabc

As pointed out, the group formed by the matrices $W(m,n)$ implies that the involved WLs
are defined as integrated along the closed curves all with one and the same base-point
$\bk_B$. We now show how two sets of WLs attributed to different base-points
are related one to another.

\begin{figure}[h]
\begin{center}
\includegraphics{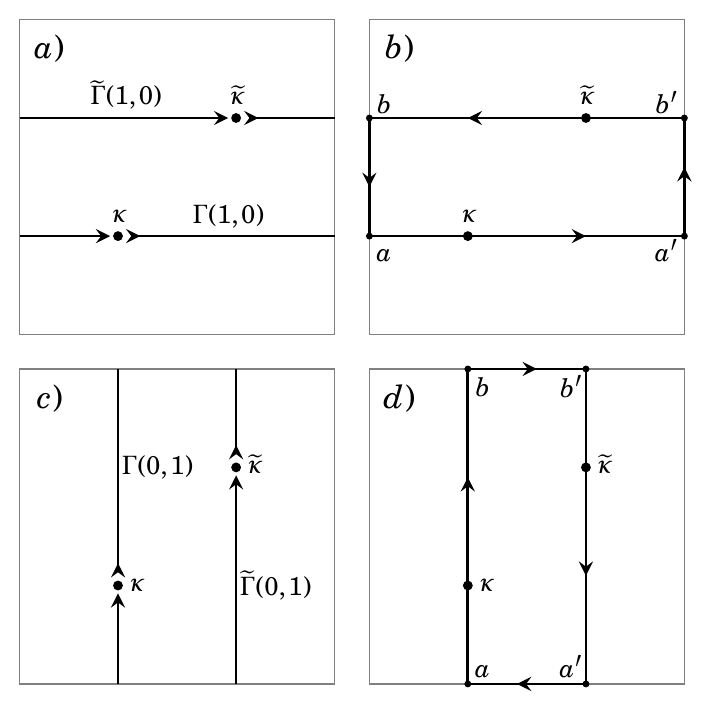}
\end{center}
\caption{{\bf Top left}: Two paths of $(1,0)$-class and different base-points $\k$ and $\tilde\k$.
{\bf Top right}: WL along $\k\to a'\to b'\to\tilde\k\to b\to a\to\k$ is trivial.
{\bf Bottom left}: Two curves of $(0,1)$-class and different base-points $\k$ and $\tilde\k$.
{\bf Bottom right}: WL along $\k\to b\to b'\to\tilde\k\to a'\to a\to\k$ is trivial.}
\end{figure}

In Fig. 6a we show two loops both belonging to $(1,0)$-class of homotopy but with different base-points $\k$ and $\tilde\k$.
For the path $\k\to a'\to b'\to\tilde\k\to b\to a\to\k$ (Fig. 6b) we have $W=\mathbb1$. From this we obtain
\be
\widetilde W(1,0)=U^\dag W(1,0)U,
\ee
where $\widetilde W(1,0)$ and $W(1,0)$ are the WLs for $\widetilde\Gamma(1,0)$ and $\Gamma(1,0)$
respectively, while $U\equiv W_{\tilde\k\to b\to a\to\k}$.

Consider now the two closed curves shown in Fig. 6c, both belonging to $(0,1)$-class but with base-points $\k$ and $\tilde\k$.
For the path $\k\to b\to b'\to\tilde\k\to a'\to a\to\k$ (Fig. 6d) we have $W=\mathbb1$ where from we find
\be
\widetilde W(0,1)=U^\dag W(0,1)U.
\ee
where the unitary matrix $U$ is the same as in (45).

Taking into account (41) we find
\be
\widetilde W(m,n)=U^\dag W(m,n)U,
\ee
{\it i.e.} the groups formed by the WLs with different base-points are unitary equivalent.

\section*{6. Four-Band Model}

As another example we consider the $4\times4$ Hamiltonian
\babc
\be
H=\lgr\ba{cc}0&t^\dag\vspace*{1mm}\\t&0\ea\rgr,
\ee
where $t$ is $2\times2$ matrix
\be
t=(a+ib)+i(\boa+i\bob)\cdot\bs,
\ee
\eabc
with $a$, $a_{1,2,3}$, $b$, $b_{1,2,3}$ being $\bk$-dependent real quantities.

The spectrum is given by
\babc
\be
E_{1,2,3,4}=(+E_+,+E_-,-E_-,-E_+)
\ee
\be
E_\pm\equiv(a^2+b^2+\boa^2+\bob^2\pm|\bh|)^{1/2}\h>0,
\ee
\be
\bh\equiv-2\hh[a\bob-b\boa+(\boa\times\bob)].
\ee
\eabc
The corresponding eigenstates are
\babc
\begin{align}
\psi_1=\frac{1}{\sqrt2}\hspace*{-1mm}\lgr\ba{c}+\hh\lambda_+\vspace*{1.5mm}\\(t\h/\hh E_+)\hh\lambda_+\ea\rgr,\\
\psi_2=\frac{1}{\sqrt2}\hspace*{-1mm}\lgr\ba{c}+\hh\lambda_-\vspace*{1.5mm}\\(t\h/\hh E_-)\hh\lambda_-\ea\rgr,\\
\psi_3=\frac{1}{\sqrt2}\hspace*{-1mm}\lgr\ba{c}-\hh\lambda_-\vspace*{1.5mm}\\(t\h/\hh E_-)\hh\lambda_-\ea\rgr,\\
\psi_4=\frac{1}{\sqrt2}\hspace*{-1mm}\lgr\ba{c}-\hh\lambda_+\vspace*{1.5mm}\\(t\h/\hh E_+)\hh\lambda_+\ea\rgr,
\end{align}
\eabc
where $\lambda_\pm$ satisfy the equations
\be
[(a^2+b^2+\boa^2+\bob^2)\mathbb1+\bh\cdot\bs]\hh\lambda_\pm=E^2_\pm\lambda_\pm.
\ee

Remark that apart the term proportional to $\mathbb1$ the matrix in the left-hand side of (51) coincides with the 2-band Hamiltonian (23).
Therefore the solutions $\lambda_\pm$ are identical to (25)
\babc
\begin{align}
\lambda_+&=\frac{1}{\sqrt{2h(h-h_3)}}\lgr\ba{c}h_1-ih_2\vspace*{1.5mm}\\+h-h_3\ea\rgr,\\
\lambda_-&=\frac{1}{\sqrt{2h(h-h_3)}}\lgr\ba{c}-h+h_3\vspace*{1.5mm}\\h_1+ih_2\ea\rgr.
\end{align}
\eabc
where $h\equiv|\bh|$ and $h_{1,2,3}$ are now given by (49c).

The Berry connection matrix is given by
\babc
\begin{align}
(A_\mu)_{11}&=(A_\mu)_{44}
=i\hh\lambda^\dag\hspace*{-1.5mm}_+\partial_\mu\lambda_+
+\frac{i\hh\lambda^\dag\hspace*{-1.5mm}_+\big[t^\dag\partial_\mu t-(\partial_\mu t^\dag)\hh t\big]\hh\lambda_+}{4E_+E_+},\\
(A_\mu)_{22}&=(A_\mu)_{33}
=i\hh\lambda^\dag\hspace*{-1.5mm}_-\partial_\mu\lambda_-
+\frac{i\hh\lambda^\dag\hspace*{-1.5mm}_-\big[t^\dag\partial_\mu t-(\partial_\mu t^\dag)\hh t\big]\hh\lambda_-}{4E_-E_-},\\
(A_\mu)_{12}&=(A_\mu)_{43}=
\frac{i}{2}\Bigg(\frac{E_-}{E_+}+1\Bigg)\lambda^\dag\hspace*{-1.5mm}_-\partial_\mu\lambda_+
+\frac{i\lambda^\dag\hspace*{-1.5mm}_-(t^\dag\partial_\mu t)\hh\lambda_+}{2E_+E_-},\\
(A_\mu)_{13}&=(A_\mu)_{42}=
\frac{i}{2}\Bigg(\frac{E_-}{E_+}-1\Bigg)\lambda^\dag\hspace*{-1.5mm}_-\partial_\mu\lambda_+
+\frac{i\lambda^\dag\hspace*{-1.5mm}_-(t^\dag\partial_\mu t)\hh\lambda_+}{2E_+E_-},\\
(A_\mu)_{14}&=\frac{i\lambda^\dag\hspace*{-1.5mm}_+\big[t^\dag\partial_\mu t-(\partial_\mu t^\dag)\hh t\big]\hh\lambda_+}{4E_+E_+},\\
(A_\mu)_{23}&=\frac{i\lambda^\dag\hspace*{-1.5mm}_-\big[t^\dag\partial_\mu t-(\partial_\mu t^\dag)\hh t\big]\h\lambda_-}{4E_-E_-},
\end{align}
\eabc
and the rest components can be obtained from $A^\dag=A$.

In the given 4-band model we have three gaps: one mid-gap occurring when $E_-(\bk)\ne0$ and two side-gaps
occurring when $E_+(\bk)\ne E_-(\bk)$ {\it i.e.} when $h(\bk)\ne0$. Singularities related to the gap closings
are explicitly presented by the factors of $E_-$ and $h$ in the denominators of (52) and (53). On top of that
we have the factor of $h-h_3$ in the denominator of (52), which may give rise to the singularity not related to
any gap closing.

As pointed out in Sect. 5 we discuss the gapped states only, since the gapless ones bring in the level crossing issues
poorly understood in the sense of the Berry (adiabatic) phases. Hence we discuss an isolated singular point $\bk_0$
(if any) where $h=h_3$. Consider the WL for $\gamma$ set by $\bk=\bk_0+\varepsilon(\cos\alpha,\sin\alpha)$
with $0\leqslant\alpha\leqslant2\pi$ and $\varepsilon\to0$. Like in the 2-band model, only the diagonal terms of
$A_\mu d\k_\mu$ survive as $\varepsilon\to0$. Consequently, the path-ordering becomes irrelevant.
Then $W_\gamma=e^{-i\Phi(\bk_0)}$ with
\be
\Phi(\bk_0)\equiv\oint A_\mu d\k_\mu=2\pi\mathbb Z\hh{\rm diag}(+1,-1,-1,+1)
\ee
where from $W_\gamma=\mathbb1$, hence the WL for an arbitrary closed path encircling the isolated point $\bk_0$
is trivial $W_\Gamma=\mathbb1$. Therefore the arguments presented in Sects. 5.1 -- 5.4 are valid also for the 4-band model,
and we come up to the same group structure of WLs revealed in the preceding sections.

\section*{7. Summary}

We have considered a tight-binding model set in terms of a hermitian matrix Hamiltonian defined on a 2D Brillouin zone and producing
multiband energy spectrum. Provided $\psi_n(\bk)$ is the $n$'th band eigenstate, the non-Abelian Berry connections
$(A_\mu)_{mn}=i\hspace*{0.3mm}\psi^\dag\hspace*{-1.4mm}_n\partial_\mu\psi_m$ are shown to be pure gauge with point-like
singularities, {\it i.e.} the curvature $F_{\mu\nu}=\partial_\mu A_\nu-\partial_\nu A_\mu+i\hspace*{0.3mm}[A_\mu,A_\nu]$
identically vanishes in the Brillouin zone except some isolated points where $F_{\mu\nu}$ exhibits singular behaviour.

Combining the singular behaviour of $F_{\mu\nu}$ with non-Abelian Stokes theorem we express the Wilson loop for a solitary
singular point in terms of the Berry phases.

We distinguish between two different kinds of singularities: one appearing in gapped states and the other in gapless states.
Gapless states imply the existence of level crossing for which the Berry phases are nor well-defined. Therefore we study the
gapped states only using the examples of 2-band and 4-band models. In those cases we show that the set of WLs attributed to
a given base-point is isomorphic to the fundamental group of the torus (Brillouin zone). Two sets of WLs with
different base-points are shown to be unitary equivalent.

\section*{Acknowledgments}

One of the authors (T.S.) acknowledges support from the World Federation of Scientists.

\begin{widetext}

\newpage

\end{widetext}

\end{document}